\documentstyle[aps,epsfig]{revtex}                         

\draft
\begin{document}
\twocolumn[\hsize\textwidth\columnwidth\hsize
     \csname @twocolumnfalse\endcsname



\title{Simulation of  Einstein-Podolsky-Rosen experiments in a local hidden
       variables model with limited efficiency and coherence.
       }

\author{W. A. Hofer}
\address{
         Department of Physics and Astronomy,
         University College, Gower Street, London WC1E 6BT, UK}

\maketitle

\begin{abstract}
We simulate correlation measurements of entangled photons
numerically. The model employed is strictly local. The correlation
is determined by its classical expression with two decisive
difference: we sum up coincidences for each pair individually, and
we include the effect of polarizer beam splitters. We analyze the
effects of decoherence, detector efficiency and polarizer
thresholds in detail. The Bell inequalities are violated in these
simulations. The violation depends crucially on the threshold of
the polarizer switches and can reach a value of 2.0 in the
limiting case. Existing experiments can be fully accounted for by
limited coherence and non-ideal detector switches. It seems thus
safe to conclude that the Bell inequalities are no suitable
criterium to decide on the nonlocality issue.
\end{abstract}


\vskip2pc]

The Einstein-Podolsky-Rosen (EPR) problem has long occupied a
central place in the understanding of quantum mechanics
\cite{epr35}. Bell's inequalities in conjunction with correlation
measurements seemed to prove that reality in microphysics is
manifestly nonlocal \cite{bell64,aspect82,weihs98}. Furthermore,
the experimental evidence seems to contradict even the notion of
an independent reality \cite{espagnat89}. Both of these features,
if true, are highly problematic. The former, because no field
propagating with a velocity exceeding c has ever been observed.
The latter, because without an independent reality there is no
guarantee that theoretical models can at all be contradicted. And
without the possibility of contradiction scientific progress
follows no clear rules.

For these reasons the EPR problem is far more important than the
experiments alone indicate. Consequently, a large amount of work
has been devoted to this field. Two years ago, the standard
reference on EPR - the book by Afriat and Selleri \cite{afriat99}
- more or less highlighted the dilemma. But in the same year
Deutsch and Hayden \cite{deutsch99} could show, by an analysis of
the information flow in such an experiment, that there is in fact
no nonlocal connection between the two measuring devices. All
information about the two angles of polarization, $\phi_{1}$ and
$\phi_{2}$, is stored locally. Even though this information cannot
locally be accessed. It is probably due to this new field of
research, quantum information theory, that the problem is even
more important today than it was ten years ago. Consequently, a
number of papers in the last two years have analyzed the paradox
from different angles, and the analysis brought two features into
focus: (i) The validity and significance of Bell's inequalities
\cite{deutsch99,sica99,adenier00}; and (ii) the relevance of a
photon's phase for the correlations
\cite{unni00,hofer00,anasto00,enk01}. From the viewpoint of
information flow a violation of Bell's inequalities is no proof of
nonlocality \cite{deutsch99}. From a formal point of view it seems
that the standard inequalities cannot be derived without violating
established notions about the measurement process. For a detailed
discussion see Sica or Adenier \cite{sica99,adenier00}. The notion
of a phase seemed initially problematic because the phase e.g. of
a wavefunction cannot be related to physical properties of a
photon. But as shown later, the phase indicates the phase of a
photon's electromagnetic field \cite{hofer00}. And it could be
established that the existence of a phase connection between the
two points of measurement, a connection which arises at the
process of emission from a common source, is sufficient to explain
correlations between two measurements in space-like separation. It
was also emphasized that measurements cannot in general be
factorized without loosing the linearity of the fields between the
two polarizers.

In this Letter we pursue a different strategy. We perform
numerical simulations of actual experiments. We sought to include
the features of the experimental situation as far as possible. For
this reason we shall give not only the results of ideal
measurements, but also measurements with limited efficiency,
limited coherence, and under the condition of dead angles of our
polarizers. It will be seen that all these effects have a bearing
on the actual data.

\begin{figure}
\begin{center}
\epsfxsize=1.0\hsize \epsfbox{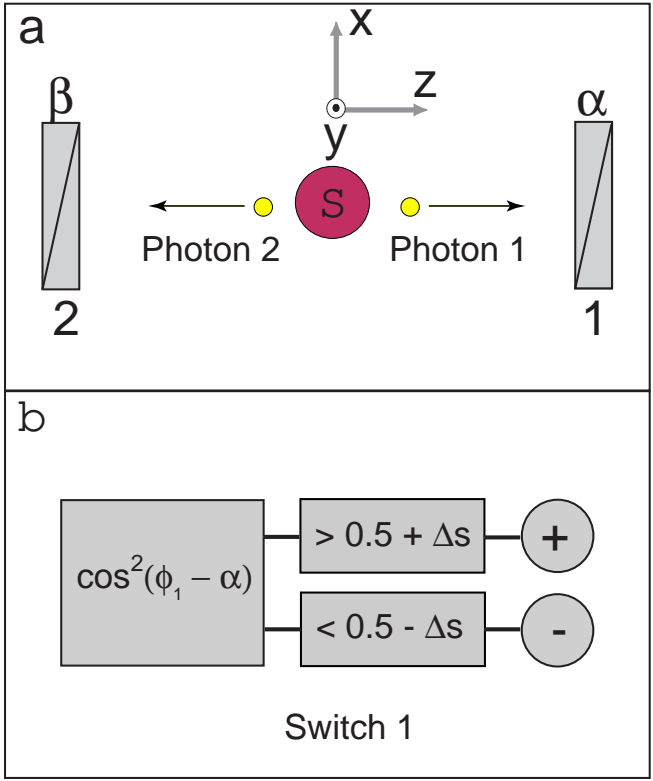}
\end{center}
\vspace{0.5 cm} \caption{
         One dimensional model of EPR type experiments. ({\bf a})
         The measuring devices (1) and (2) are in opposite directions
         from the photon source S. The polarizers are set to the angles
         $\alpha$ and $\beta$, respectively.  ({\bf b}) The switches at both
         stations measure the polarization and, depending on the angle,
         either transmit 1 or 0 to the computer. Note that the dead angle
         of the polarizer is simulated by a threshold $\Delta$s of the
         switch.
        }
\label{fig001}
\end{figure}

{\em Setup.} - The setup of the experiment is shown in Fig.
\ref{fig001} (a). A source between two polarizers emits a pair of
photons along the $z$ axis. Two polarizers, at the positions
$L_{1}$ and $L_{2}$, respectively, measure the angle of
polarization. The angle of polarizer one is varied by a half
cycle, $\pi$, during the experiment. At every position of the
polarizer a set of 1000 photon pairs is emitted and measured. The
switch of the polarizer is shown in Fig. \ref{fig001} (b). If
$\cos^2(\phi_{1} - \alpha)$ is larger than $0.5 + \Delta s$, then
the event is recorded as a transmission (+). If it is less than
$0.5 - \Delta s$, it is recorded as an adsorption (-). No action
is taken for values between these two boundaries. The same switch
is applied to both measurement devices. The threshold $\Delta s$
formalizes three separate features of the experiments: (i) The
transmission characteristics of the polarizer beam splitters; (ii)
the threshold of the photodetectors; and (iii) the electronic
evaluation of events, since double counts ($+$ {\em and} $-$) are
excluded.

A single simulation run starts with the initialization of the
random number generator \cite{random}. The generator is
initialized only once, at the beginning of a full simulation
cycle. The random number is mapped onto the initial phase from 0
to $2 \pi$ of the photon pair. Simulations are generally made with
a phase difference of $\pi/2$ between the angles of polarization
of photon one and photon two. After covering the distance to
polarizer one and two the photons are measured. We assumed,
without lack of generality, that both distances are integer
multiples of the wavelength. After a single pair has been
measured, we record the coincidences \mbox{(++,+-,-+,--)}. The
procedure is repeated for all 1000 pairs, then the polarization
angle of device one is changed by $\pi/100$. A run ends, when all
1000 pairs at the final position of polarizer one have been
measured ($\pi$). In all figures we only plot the coincidences
$N_{++}$.

We accounted for limited efficiency and decoherence in the
following way. Limited efficiency means that not all pairs emitted
are actually measured. In this case we simply did not evaluate all
pairs, depending on the efficiency of the setup. 50 \% efficiency,
for example, means that only every second pair is actually
recorded. To simulate decoherence we created an independent random
input for a certain fraction of a half cycle of $\pi$. 100 \%
decoherence here means that half a wavelength of the photon's
optical path is random. This translates into a polarization angle
random in the interval [0,$\pi$]. Both effects reduce the maximum
of the output measured, but it will be seen presently that they
have very different effects.

\begin{figure}
\begin{center}
\epsfxsize=1.0\hsize \epsfbox{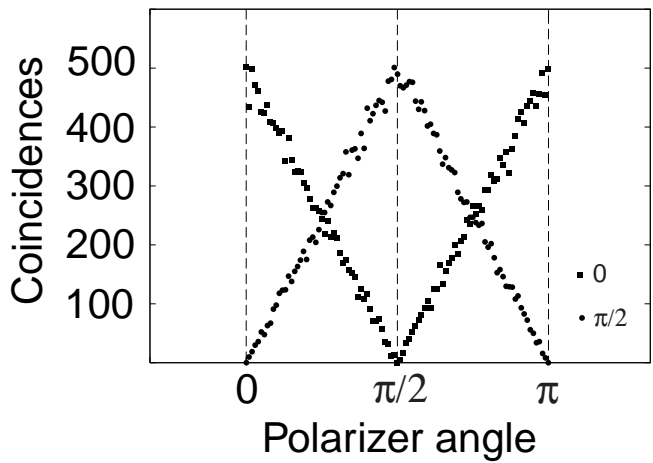}
\end{center}
\vspace{0.5 cm} \caption{
         Ideal measurement of correlations ($N_{++}$). Two simulations were
         performed with a difference of 0 (full squares) and
         $\pi/2$ (full circles) of the polarization angle at the
         origin. Neither of these curves is equal to the
         theoretical prediction, instead we obtain an angular
         shape for the correlations.
        }
\label{fig002}
\end{figure}

{\em Ideal measurements.} - Initially we simulated an ideal
measurement. The efficiency in this simulation is 100\%, the
fields of the two photons are fully coherent throughout the
distance between the two polarizers, and the experimental devices
are supposed to have ideal characteristics. The result of this
simulation is shown in Fig. \ref{fig002}. We did two separate
simulations, one with a polarization difference between the two
photons of 0 (full squares), the other with $\pi/2$ (full
circles). It can be seen that neither of the simulations comes
close to the theoretical prediction of a $sin^2(\beta - \alpha)$.
Instead, the curves representing ideal measurements would be of
angular shape. However, the maximum of the correlation ($N_{0}/2$,
where $N_{0}$ is the number of pairs) and the minimum (zero) are
exactly obtained in the extreme cases.

It should be noted that the results given in these plots reflect
the ''classical'' formulation of a coincidence, given by the
equation \cite{furry36}:

\begin{equation}
P(\alpha,\beta) = \int d \lambda \cos^2(\lambda - \alpha) \cos^2
(\lambda - \beta)
\end{equation}

with two decisive differences: first, the summation is performed
over single pairs, as in the actual experiments, rather than over
the two polarizers separately. The latter procedure, formalized in
the given integral, includes not only photons of one pair, but
also sums up contributions of different pairs. And second, we
accounted for the digital output of the polarization devices. The
integral is no suitable representation of the digital results in
current experiments. If, for example, $\lambda = \phi_{1}$ and
$\phi_{2} = \phi_{1} + \pi/2$, then for $\alpha = \beta = 0$ the
integral yields $N_{0}/8$. But the actual count, under the
condition that $\cos^2 \phi_{1} > 0.5$ and $\cos^2 (\phi_{1} +
\pi/2) > 0.5$ is zero. The digitalization, necessary to obtain
formal agreement with spin measurements in quantum mechanics, is
usually achieved by means of a polarizer beam splitter
\cite{weihs98}.

{\em Dead angles.} - In our simulations we find that the curves
obtained are not very sensitive to the threshold of the polarizer
switches. We have performed simulations where $\Delta s$ was
varied from 0.00 to 0.20. Apart from a reduction of the absolute
yield the increase of the threshold only affects the width of the
minimum at the ultimate angles. This effect is equal to a
retardation of the onset of the correlation function at its
minimum position. The threshold therefore does not change the
functional form of the correlations.

\begin{figure}
\begin{center}
\epsfxsize=1.0\hsize \epsfbox{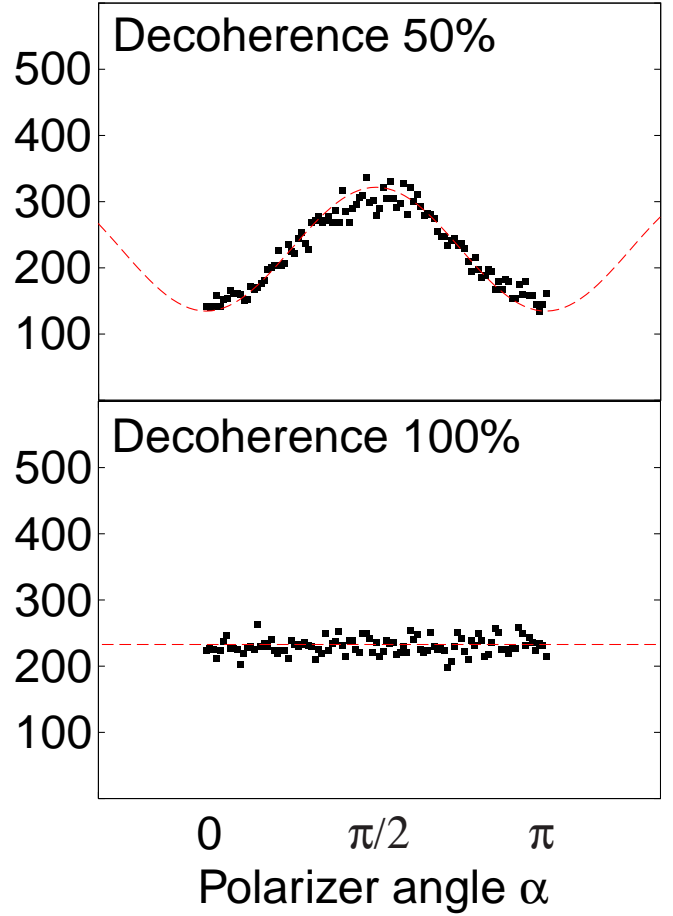}
\end{center}
\vspace{0.5 cm} \caption{
         Dependence of coincidence rates on the efficiency of detection.
         The shape of the distribution is similar to the ideal distribution,
         but the statistical spread is considerably larger. We include the
         $\sin^2(\alpha)$ function for reasons of comparison. This function
         has not been actually fitted to simulated data.
        }
\label{fig003}
\end{figure}

{\em Efficiency.} - The detection of photons is one of the
problems experimenters are still confronted with. The efficiency
is in fact so low (less than 10\% \cite{weihs98}), that the
correlations found in Aspect's measurements \cite{aspect82} were
disputed on the grounds of an ''efficiency loophole''. In our
simulations such a conclusion would not be justified. Even though
the shape of the curve changes somewhat and the statistical spread
is dramatically increased in the low efficiency range, the maxima
and minima are still clearly distinguishable. The minimum,
moreover, remains zero. We included a dead angle of detectors by a
threshold of 0.05. The same threshold will be used in subsequent
simulations. The coincidence rates due to detector efficiency
varying from 50\% to 10\% are shown in Fig. \ref{fig003}. From
this result we conclude that efficiency is not the decisive issue
to estimate the relevance of an experiment.

{\em Decoherence.} - The polarizers in current measurements are
more than 400 m apart \cite{weihs98}. Furthermore, there is no
vibration damping or cooling to very low temperatures involved in
such a measurement. This feature of the measurements is bound to
cause random motion of system components. From surface science the
range of motion without damping can be estimated, it should be for
an isolated surface no less than a few nanometer or more than one
percent of the photon's wavelength. Considering that we deal with
three coupled components and optical paths in between it seems
safe to increase this estimate by one order of magnitude. In this
case we have to include random motion of our system in the range
of about 5-10\% of the wavelength. This translates, in our
simulation, into a rate of decoherence of 10-20\% (100\% means
that half a wavelength of the photon's optical path is random).

\begin{figure}
\begin{center}
\epsfxsize=1.0\hsize \epsfbox{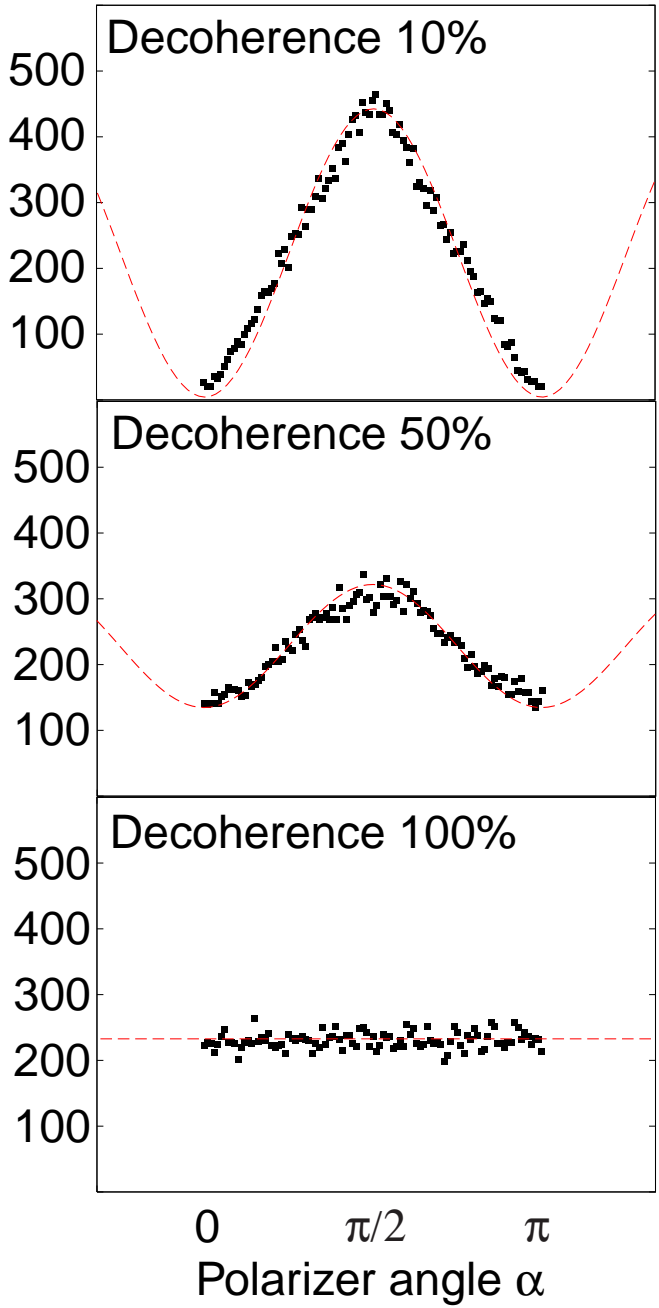}
\end{center}
\vspace{0.5 cm} \caption{
         Dependence of coincidence rates on the decoherence of photon beams.
         Due to decoherence the distribution becomes more sinusoidal. In the
         limit of full decoherence we obtain uncorrelated measurements.
        }
\label{fig004}
\end{figure}

Simulation with a decoherence rate of 10\%, 50\% and 100\% are
shown in Fig. \ref{fig004}. The interesting feature of decoherence
is that it renders the resulting distribution more sinusoidal than
the correlations of an ideal measurement. The fully decoherent
simulation proves that correlations are independent of the setting
as required, but it also shows the noise due to a random
distribution of the initial phase of the coupled system. In
practice all effects analyzed will to a greater or lesser extent
be present in any single measurement.

\begin{figure}
\begin{center}
\epsfxsize=1.0\hsize \epsfbox{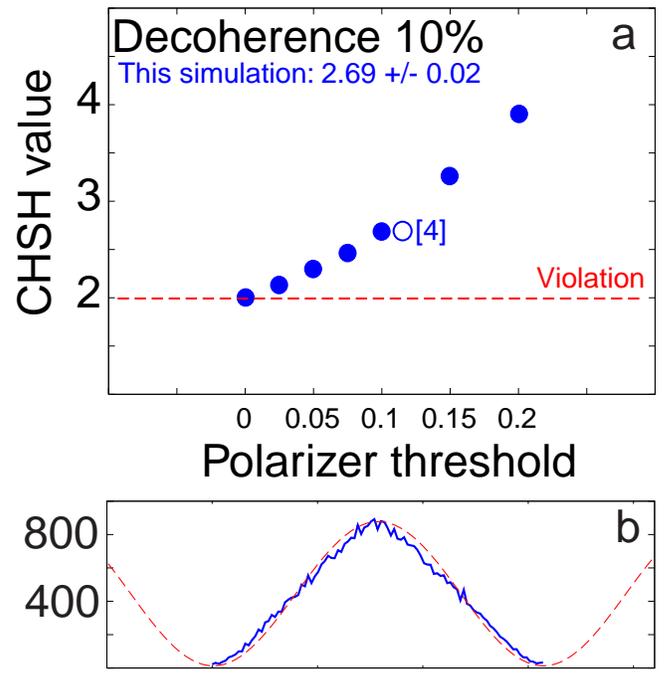}
\end{center}
\vspace{0.5 cm} \caption{
         Violation of Bell's inequalities depending on the
         threshold of the polarizer switches. ({\bf a})
         The inequality is violated in all cases where the
         threshold is not zero (full circles).
         We obtain a maximum CHSH value of
         3.9 (threshold 0.2). The experiments of Weihs et al.
         indicate a threshold of 0.1 (empty circle).
         ({\bf b}) Simulation of EPR experiments with a
         decoherence of 10\% and a threshold of 0.1. It can be
         seen that the distribution differs only insignificantly
         from the square of a sinus.
        }
\label{fig005}
\end{figure}

{\em Bell violations.} - Finally, we demonstrate the influence of
the polarizer threshold on the violation of Bell's inequalities.
To this end we simulate the counts at four selected angular
positions of the polarizers $\alpha$ and $\beta$ (0$^{\circ}$,
45$^{\circ}$, 22.5$^{\circ}$, 67.5$^{\circ}$). These positions
yield the maximum violation of Bell's inequalities in the standard
framework. We performed the simulations for varying threshold
values from 0.0 (no threshold) to 0.2 (nearly half the photons
remain undetected). For every setting we performed 10 separate
runs, each with 10000 pairs of photons, the efficiency of the
detectors was assumed to be 100\%. Fig. \ref{fig005} gives the
result of our simulation. The violation (computed according to the
version of Clauser et al. (CHSH) \cite{clauser69}) increases with
increasing threshold. Furthermore, the limit of violation is close
to 2.0 (CHSH value of 3.90) in the final setting. The actual
threshold in the experiments can be estimated from the visibility
of the correlation function. For decoherence rates of 10 - 20 \%,
as in the experiments, we can only obtain agreement with reported
values (97\% visibility \cite{weihs98}), if we increase the
threshold $\Delta s$ to more than 0.10. Our simulations indicate
that the violation depends crucially on the threshold. The maximum
violation can reach a value of as much as 2.0 (the limiting case).
It seems thus safe to conclude that the Bell inequalities are no
suitable criterium to decide on the nonlocality issue.

{\em Summary.}- In summary we have presented a numerical
simulation of EPR experiments under the assumption of strict
locality and analyzed the effects of polarizer thresholds, limited
efficiency, and decoherence in detail. We could show that Bell's
inequalities are violated in these simulation, and that the
violation depends crucially on the threshold of the polarizer
switches. The limit of violation in this model is about 2.0 (CHSH
value of 4.0). We found that existing measurements can be fully
accounted for by limited coherence and non-ideal polarizer
switches.

{\em Acknowledgements.} - Discussions about the EPR problem over
the last year proved very helpful. I am especially indebted to A.
Fisher, J. Gavartin, R. Stadler, and B. Wolkow. Computing
facilities at the HiPerSPACE center were funded by the Higher
Education Funding Council for England.

%

\newpage

%

\end{document}